\documentclass[conference]{IEEEtran}

\usepackage{amsmath}

\usepackage{indentfirst}

\usepackage{graphicx}
\usepackage{courier}
\usepackage{float}
\usepackage{subfig}

\usepackage{flexisym} 
\usepackage{tabularx}
\usepackage{pdfcomment}
\usepackage{algpseudocode}
\usepackage{algorithm}
\usepackage{varwidth}

\usepackage{url}

\usepackage{pdfcomment}

\usepackage{multirow}

\DeclareMathOperator*{\argmin}{arg\,min}

\algnewcommand{\algorithmicgoto}{\textbf{go to}}%
\algnewcommand{\Goto}[1]{\algorithmicgoto~\ref{#1}}%

\begin{document}
%
\title{Budget Constrained Execution of Multiple Bag-of-Tasks Applications on the Cloud}

\author{\IEEEauthorblockN{Long Thai, Blesson Varghese and Adam Barker}
\IEEEauthorblockA{School of Computer Science, University of St Andrews, Fife, UK\\
Email: \{ltt2, varghese, adam.barker \}@st-andrews.ac.uk}
}

\maketitle

\begin{abstract}
Optimising the execution of Bag-of-Tasks (BoT) applications on the cloud is a hard problem due to the trade-offs between performance and monetary cost. The problem can be further complicated when multiple BoT applications need to be executed. In this paper, we propose and implement a heuristic algorithm that schedules tasks of multiple applications onto different cloud virtual machines in order to maximise performance while satisfying a given budget constraint. Current approaches are limited in task scheduling since they place a limit on the number of cloud resources that can be employed by the applications. However, in the proposed algorithm there are no such limits, and in comparison with other approaches, the algorithm on average achieves an improved performance of 10\%. The experimental results also highlight that the algorithm yields consistent performance even with low budget constraints which cannot be achieved by competing approaches. 
\end{abstract}

\IEEEpeerreviewmaketitle

\section{Introduction}
\label{introduction}
Bag-of-Tasks (BoT) is defined as a collection of independent and identical tasks, which can be executed by the same application but in any order. It is possible to split a BoT into sub-BoTs, each of which is assigned to one separate machine for execution. As a result, BoT applications are usually executed in a distributed environment, for instance, they account for more than 75\% of Grid computing workloads \cite{Iosup:2011:IEEEIC}.

With the advent of cloud computing  \cite{hotcloud} distributed computing resources ranging from basic to compute optimised, or memory optimised machines are available on a pay-as-you pricing scheme. Cloud computing therefore offers a cost-effective solution to execute BoT applications, in which a user is free to choose the type and quantity of resources required for her application. 


A key challenge when executing BoT applications on the cloud in order to achieve maximum performance is the trade-off between decreasing the time it takes to execute individual tasks and increasing the number of tasks executed at the same time. Using high-performing (but expensive) machines can reduce the time to execute an individual task. On the other hand, a larger collection of cheaper machines will maximise execution parallelism. An additional challenge is encountered when a user needs to execute multiple BoT applications at the same time, as each application will differ in performance on the same type of machine. For example, tasks of a CPU intensive application will perform best on a compute optimised machine; a memory optimised machine may not be best suited.

In this paper, a heuristic algorithm which considers the diversity in cost, machine types and application performance is proposed to solve the problem of executing multiple BoT applications on the cloud given a user's budget constraint. The algorithm efficiently assigns tasks to cloud machines of different types such that the budget constraint is not violated while minimising the execution time. The algorithm is evaluated against existing approaches and achieves better performance for a given budget.

The research contributions of this paper are as follows: (i) a mathematical model of the problem of executing BoT on the cloud while taking into account a budget constraint, ii) the development and implementation of a heuristic algorithm that aims to maximise the performance of a BoT on the cloud while satisfying the given constraint, and iii) an evaluation which compares the proposed algorithm with other approaches.

The remainder of this paper is organised as follows. Section \ref{relatedwork} considers research related to that presented in this paper. Section \ref{model} presents a mathematical model of the problem. Section \ref{algorithm} proposes the algorithm for executing multiple BoT applications. Section \ref{evaluation} evaluates the proposed algorithm. Section 6 concludes this paper by considering future work.

\section{Related Work}
\label{relatedwork}
One popular framework for executing BoT is BOINC \cite{BOINC}, which distributes tasks to resources whose computation is volunteered from around the world.


The MyGrid \cite{Cirne:2003:ICPP} framework facilitates the execution of a BoT on the grid by minimising the execution time. This is achieved by replicating and assigning unfinished tasks to idle resources. Task scheduling algorithms have been previously investigated \cite{Maheswaran:1999:DMS}. The location of input data can be taken into account to reduce the execution time of BoT and improve the Quality-of-Service (QoS) \cite{Ranganathan:2002:DCD, Weng:2005:HSB}. Independent file-sharing tasks can be executed on the grid efficiently by preventing the bottleneck of all machines executing the tasks requiring to download data from the centralised server \cite{Kaya:TransPDS:2006}. Scheduling algorithms in which each task requires data distributed at multiple sources and satisfies both deadline and budget constraints have been considered \cite{Venugopal:2005:ICA3PP}.

Executing multiple BoT applications is also widely investigated by researchers. There are decentralised approaches to increase the throughput and fairness of the execution \cite{Bertin:2008:Grid}. Another strategy allows multiple tasks to be executed concurrently on the same machine without severely affecting performance \cite{Benoit:TransComp:2010}. An evaluation of different strategies to execute multiple BoT applications on the Grid is considered by Anglano and Canonico\cite{Anglano:2008:IPDPS}.


However, those researches in Grid computing may not be applicable to cloud environment as their resources are already available (i.e. machines are already running) and usually free of charge. On the other hand, a cloud user has to decide (and pay for) the type and the number of resources required \textbf{before} the actual execution. Furthermore, the applicability of Grid computing is not as wide as cloud research since those platforms are mostly accessible to organisations that can afford to invest into the infrastructure and maintain it.

Recently, researchers have started to focus on employing the cloud for executing BoT. For example, statistical approaches to schedule BoT on the cloud given a budget constraint \cite{Oprescu:2010:CloudCom} and approaches to cost-effectively execute BoT on multiple clouds \cite{Farahabady:2012:PDCAT} are recent efforts. A comparison of scheduling algorithms for executing multiple BoTs on the cloud has been investigated \cite{Gutierrez-Garcia:2013:FHA}. Mao et al. propose a approach to scale Cloud resource based on deadline and budget constraints using constraint programming \cite{Mao:2010:GRID}. In our previous work \cite{Thai:2014:CloudCom}, we investigate the execution of a Bag-of-Distributed-Tasks (BoDT) application, in which each task required data from a globally distributed source. Hence, the BoDT application is split into multiple BoT applications, each of which only contains tasks from one data source. Due to the geographical and network distance, the task execution, which includes downloading input data, of tasks from different applications, i.e. data source, can be different. With the same amount of money, a user can obtain either a small number of high performance but expensive machines or many low performance but cheap ones. The trade-off between application makespan and execution parallelism is investigated in \cite{Appuswamy:2013:SVS}.

In comparison with \cite{Oprescu:2010:CloudCom,Farahabady:2012:PDCAT,Thai:2014:CloudCom} which make an assumption that there is a limit to the number of cloud resources, our paper allows a user to acquire as many resources allowed by the budget. Moreover, it performs not only resource provisioning \cite{Oprescu:2010:CloudCom,Mao:2010:GRID}, but also task assignment for multiple applications to cloud resources. Even though task assignment is more complicated due to the high number of tasks, it offers a better flexibility and is more suitable for cases when the execution time of each task is not similar due to additional factor such as their data size.

\section{Problem Modelling}
\label{model}
In this section, the problem of executing multiple BoT applications on the cloud with budget constraint is modelled.

\subsection{System Model}
Let $M$ be the number of applications and the set of application be $A =\{A_1...A_M\}$. Each application is a collection of the same type of tasks denoted as $A_i = \{t_{i, 1}...t_{i, |A_i|}\}$.

Let $T = \bigcup_{A_i \in A}{A_i} = \{t_1, t_2, ... t_{\sum_{A_i}{|A_i|}}\}$ be the list of tasks, thus $|T| = \sum_{A_i \in A}{|A_i|}$. Each task belongs to one application ($\forall t \in T: \exists ! A_j \in A$), such that $t_i \in A_j$. For any given task $t$, its application can be found as $A_t$.

Each $t \in T$ is measured by $size_t$, which is used to compare each task \textbf{of the same application}. The size of a task can be the actual size of its input data or any parameter related to its complexity; for example, the number of training iteration for a machine learning application. The value of $size_t$ determines the time taken by the application to run on similar hardware; more execution time is required for a larger value.

Let $IT = \{it_1...it_N\}$ be the set of $N$ instance types offered by the cloud providers. The cost per hour of an instance is denoted as $c_{it}$.

The performance of each type of instance changes from one application to another since there are multiple applications. Let $P$ be the performance matrix of size $N \times M$. $P_{i, j}$ is the time in seconds taken by a instance type $it_i$ to process one unit of size of a task of an application $A_j$. For each instance type $it_i \in IT$, its performance is the vector $P_{it_i} = P_i = \{ P_{it_i, A_1} ... P_{it_i, A_M} \}$ corresponding to all applications.

In order to acquire the performance between instance types and applications, we suggest to perform some test runs as, to the best of our knowledge, there is not yet any research in predicting application's performance on different types of machine.

The execution time of a task $t$ using instance type $it$ is $exec_{it, t} = P_{it, A_t} \times size_t$. Thus, the execution time of the collection of $T$ on $it$ can be calculated as $exec_{it, T} = \sum_{t \in T}{exec_{it, t}}$.


We assume in this model that there is no pair of instance types that have the same performance and cost. So in the model it is possible to have multiple instances with the same either performance or cost.
\begin{equation} \label{eq:unique_type}
	P_{it_i} = P_{it_j} \wedge c_{it_i} = c_{it_j} \iff it_i = it_j
\end{equation}

The system in which multiple applications must be executed on the cloud consisting different VM types can be represented as $(A, IT)$.

\subsection{Problem Model}

The execution plan can be represented as the list of VMs, each of which is created from one instance type and has the list of assigned tasks.

So, assume that $VM = \{vm_1...\}$ is the execution plan in which each $vm \in VM$ is created based on one instance type $it \in IT$. For $vm \in VM$, $it_{vm}$ denotes the type of $vm$. Additionally, $VM_{it}$ be the list of VMs created from the same instance type $it$.

Let $T_{vm}$ be the list of tasks assigned to $vm \in VM$. The time to execute a task $t$ that is assigned to $vm$ is:
\begin{equation}
	exec_{vm, t} = exec_{it_{vm}, t} = P_{it_{vm}, A_t} \times size_t
\end{equation}

The following constraint must be satisfied for all tasks to be executed:
\begin{equation}
	\bigcup_{vm \in VM}{T_{vm}} = T
\end{equation}

Moreover, one task cannot be assigned to multiple VMs and this condition is represented as
\begin{equation}
	T_{vm_i} \bigcap T_{vm_j} = \emptyset \text{ if $vm_i \neq vm_j$}
\end{equation}

A start up time is required to boot a VM into a usable state and this overhead is denoted as $o$. The overhead is paid for by the user although a task cannot be executed on the VM during start up. 

The execution time of $vm \in VM$ is the sum of the time taken to execute all tasks assigned on the VM and the time for start up denoted as
\begin{equation}
	exec_{vm} = o + \sum_{t \in T_{vm}}{exec_{vm, t}}
\end{equation}

We assume that each VM is charged by hour, and hence, if only a small fraction of the hour is utilised, then the user still has to pay for the entire hour. The cost of running a $vm \in VM$ is
\begin{equation}
	cost_{vm} = \lceil \frac{exec_{vm}}{3600} \rceil \times c_{pt}
\end{equation}

The overall time to complete all tasks is the execution time of the slowest VM (all VMs execute tasks in parallel) and is denoted as
\begin{equation} \label{eq:exec}
	exec = \max_{vm \in VM}{exec_{vm}}
\end{equation}

The total cost to execute all tasks is the sum of the costs of all VMs which is
\begin{equation}\label{eq:model:budget}
	cost = \sum_{vm \in VM}{cost_{vm}}
\end{equation}

If $B$ denotes the budget constraint for the amount of money that can be spent for executing $T$ on the cloud, then 
\begin{equation}
	cost \leq B
\end{equation}

In this research, the performance of BoTs on the cloud is maximised by determining $VM$, referred to as an \textbf{execution plan}, which contains a set of VMs and the assignment of tasks onto the VMs, so that the overall execution time, $exec$, is minimised while satisfying the budget constraint.

\section{Heuristic Algorithm}
\label{algorithm}
This section presents the algorithm used to solve the problem of executing multiple BoT applications on the Cloud. The main steps of the algorithm include, creating VMs, assigning tasks to VMs, balancing tasks between VMs, generating an initial plan based on local performance, adding more VMs based on the user's budget, keeping VMs' execution times under one hour, replacing expensive VMs by cheaper ones and finding an execution plan.

Our approach to address the problem consists of algorithms which are presented in Sections \ref{sec:al:assign} to \ref{sec:al:replace}. Section \ref{sec:al:find} presents the complete approach.


\subsection{Assign Tasks To VMs} \label{sec:al:assign}

Function $ASSIGN$ aims to assigns a list of tasks to a given list of VMs. For each task, a receiving VM is selected based on three criteria: i) the cost of a VM should not increase if a task is executed in it, moreover, a receiving VM should ii) require the least time to execute a task and iii) has the lowest execution time in comparison to other VMs.

After the assignments there may be VMs without any assigned tasks, since their instance types do not have the best performance for any task.

\subsection{Balance Tasks Between VMs} \label{sec:al:balance}

When tasks are assigned to VMs of different types, it is possible to have one VM with a higher execution time than the others. As shown by Equation \ref{eq:exec}, this will affect the overall execution time. Hence, it is necessary for tasks to be evenly distributed among all VMs so that their execution can be completed nearly at the same time.  This process is performed by the function $BALANCE$ which moves tasks from VMs with highest execution times to other ones as long as the overall execution time does not increase.

\subsection{Create Initial Plan by Selecting Instance Type with Best Performance for each Application} \label{sec:al:initial}

The best instance type of an application is the one whose cost is lower than the given budget and maximises performance of an application. If there are multiple instance types that maximise application performance, then the cheapest one is selected: $it^b_{A_i} = \argmin_{it \in IT}{(P_{it, A_i}, c_it)}$.

In the initial plan generated by function $INITIAL$, the tasks are assigned to the best instance type. In other words, an application's tasks are assigned to the number of VMs of the same instance type.

For each application, the whole budget is used to hire VMs of its best instance type: $num = \lfloor \frac{B}{it^b_{A_i}} \rfloor$. As there are many applications, the budget is likely to be violated.

\subsection{Reduce cost} \label{sec:al:reduce}

As an initial plan is highly likely to violate the budget constraint, the next step, therefore, is to reduce the overall cost until the budget constraint is satisfied.

Moving task can potentially increase the cost if it results in an additional hour added for the receiving VM. So, the goal of the cost reduction process is to \textbf{completely remove a number of VMs by moving all of their tasks to other VMs} without increasing the overall cost.

The cost reduction is performed using function $REDUCE$ which tries to move \textbf{all} tasks from one VM with lowest execution time to others. The function has two modes, \textbf{local mode} only allows tasks to be moved to VMs of the same type of an initial VM, while \textbf{global mode} allows tasks to be moved to VM of any type. In order to keep task's execution time as low as possible, the function tries to move tasks to VMs whose require least time to execute them.

\subsection{Add More VMs based on Budget} \label{sec:al:add_more}
Until this stage, only the best performing VMs are used. Based on the \textbf{remaining budget}, a few more VMs can be added to increase the execution concurrency which results in lower execution time even though they are not best performing.

Function $ADD$ aims to add the most number of VMs based on the remaining budget $B_r = B - cost$. The instance type of the added VMs is the cheapest one with the lowest execution time for all tasks. By assuming that each of them would not be executed for more than one hour, it is possible to calculate a cost for a new VM, and the function keeps added new VMs until there is not enough money to add any more.

\subsection{Keep VM's Execution in One Hour} \label{sec:al:keep}

As cloud VMs are usually charged by the hour; running a VM for two hours will be similar in cost to running two VMs of the same type in parallel for one hour. Hence, we introduce function $SPLIT$ which keeps assigning tasks from a VM whose execution time is more than one hour to two VMs with the same instance type as long as the budget constraint is not violated and overall execution time decreases.

\subsection{Replace Expensive VMs by Cheaper Ones} \label{sec:al:replace}

Sometimes, it is cost-effective to use a large numbers of cheaper and moderately performing VMs than fewer expensive and high-performing VMs. For example, assuming there are two instance types $IT = \{ it_1, it_2 \}$ and one application with $10$ tasks of size 1: $A = \{ A_1 \}$. The cost and performance of $it_1$ are $\$2$ and $\$8$, which means a VM of instance $it_1$ costs $\$2$ per hour and takes $8$ seconds to execute one task of $A_1$. Similarly, the cost and performance of $it_2$ are $\$1$ and $\$10$. With the budget $B = \$2$, it is possible to have one VM of type $it_1$ and takes $8 \times 10 = 80$ seconds to execute all ten tasks of $A_1$. Alternatively, with the same budget, two VMs of type $it_2$ can be deployed. As tasks are evenly distributed to both VMs, each VM executes five tasks and takes $10 \times 5 = 50$ seconds to complete execution. The execution when two VMs of instance $it_2$ are employed is $50$ seconds. In this case, two VMs of type $it_2$ perform better.

Function $REPLACE$ aims to replace expensive VMs with cheaper ones in order to increase the cost-effectiveness of the execution. First of all, it selects the certain number of VMs and find their cost. Then, it calculates how many VMs of the \textbf{cheaper} instance type are affordable based on the cost and the remaining budget (if there is any). For simplification, only one instance type is considered of the time, which means the set of VMs has the same instance type. All tasks from the selected VMs are assigned to the set of new and cheaper VMs. After assignment, if the budget is still satisfied and the overall execution time is reduced, the selected VMs are officially replaced.

\subsection{Find an Execution Plan based on the Given Budget Constraint} \label{sec:al:find}

Algorithm \ref{al:find} is used to find the execution plan based on the given budget constraint using all functions introduces in the previous sections. First of all, the $INITIAL$ function is called to create an initial plan, in which all tasks are assigned to VMs of their best instance types possible, which are then locally reduced (Lines \ref{al:find:init}, \ref{al:find:assign} and \ref{al:find:local_reduce}).

For future comparison, the current plan, cost and execution time are stored (Lines \ref{al:find:store_plan}, \ref{al:find:store_cost} and \ref{al:find:store_exec}).

After that, the current plan is globally reduced, in which tasks can be moved to all VMs except the one which is selected to be removed. (Line \ref{al:find:global_reduce}). Additional VMs can be added if it is allowed by the remaining budget (Line \ref{al:find:add}) and tasks are balanced between all VMs (Line \ref{al:find:balance}). Then, we try to keep the execution to all VMs under one hour (Line \ref{al:find:split}). As it is not guaranteed that the current execution plan satisfies the budget constrain, the greater value between the real one and the current cost of the execution is used as an temporary budget for $REPLACE$ function, which tries to replace expensive VMs which more cheaper ones (Line \ref{al:find:replace}).

The Algorithm is an iterative process which tries to optimise the execution plan by reducing its cost and execution time. Hence, if the current plan is better than the previous one, i.e. the execution time of the cost are reduced, the iteration continues (Line \ref{al:find:continue}). Otherwise, if there is no improvement in term of cost and execution time, the plan is returned (Line \ref{al:find:terminate}).

\begin{algorithm}
	\caption{Find}
	\label{al:find}
	\begin{algorithmic}[1]
		\Function{DO\_ASSIGNMENT}{$T, IT, B$}
			\State $VM \gets INITIAL(A_T, IT, B)$														\label{al:find:init}
			\State $VM \gets ASSIGN(T, VM)$																\label{al:find:assign}
			\State $VM \gets REDUCE(VM', B, \emptyset, TRUE)$											\label{al:find:local_reduce}
			\State $cost' \gets MAX\_NUMBER$															\label{al:find:store_cost}
			\State $exec' \gets MAX\_NUMBER$															\label{al:find:store_exec}
			\State $VM' \gets VM$																		\label{al:find:store_plan}
			\Loop																						\label{al:find:start_loop}
				\State $VM \gets REDUCE(VM', B, \emptyset, FALSE)$										\label{al:find:global_reduce}
				\State $VM \gets ADD(IT, VM, B - cost)$													\label{al:find:add}
				\State $VM \gets BALANCE(VM)$															\label{al:find:balance}
				\State $VM \gets KEEP(VM)$																\label{al:find:split}
				\State $VM \gets REPLACE(IT, VM, \max{B, cost}, 1)$										\label{al:find:replace}
				\If{$cost < cost' \vee exec < exec'$}													\label{al:find:continue}
					\State $cost' \gets cost$
					\State $exec' \gets exec$
					\State $VM' \gets VM$
				\Else
					\State \Return $VM'$																\label{al:find:terminate}
				\EndIf
			\EndLoop
		\EndFunction
	\end{algorithmic}
\end{algorithm}

\section{Evaluation}
\label{evaluation}
This section evaluates our approach by comparing its performance with two approaches.

\subsection{Approaches for Comparison}
The approaches used for comparing our algorithm are as follows:

\subsubsection{Minimising Individual Task Execution Time (MI) Approach} this approach aims to minimise the execution time of any individual task by selecting the instance type which has the best performance among all tasks. It can be easily performed by invoking Algorithm $ADD$ with full budget.

\subsubsection{Maximising Parallelism (MP) Approach} in this approach, the cheapest instance type is selected so that the maximum number of VMs can be purchased based on the given budget $it^c = \argmin_{it \in IT}{(c_{it})}$.

\subsection{Environment Setup}

We built a simulation framework using Scala to evaluate the heuristic algorithm. The framework models multiple instance types of a cloud with different performance and varying costs as shown in Table \ref{tab:its}; this is input to the simulation. In cloud environment, those inputs can be obtained by sampling the applications, i.e. running the small amount of their tasks, on VMs of difference instance types. The framework then uses Algorithm \ref{al:find} to generate an execution plan. This plan is then executed for obtaining the overall cost and time.

\subsubsection{Applications} Three application $A_1, A_2, A_3$ were considered in the experiments. The first one used the same amount of compute and memory resources and the other two were CPU and memory intensive applications. Each application consisted of 250 tasks whose side are equally distributed from 1 to 5.

\subsubsection{Instance Types} We assumed that there were four instances types $it_1, it_2, it_3, it_4$. The first one was very cheap and had poor performance for all applications. The second one was a general instance type which provided the balance between compute and memory. The third and forth ones were compute and memory optimised instance types which were most suitable for CPU and memory intensive applications, respectively. The last three instance types had the same cost which was twice in comparison to the first one.

\subsubsection{Cost and Performance} The description, cost and performance of each instance type is presented in Table \ref{tab:its}. It can be seen that even though the last three instance types had the same price, they performed differently.

\begin{table}[h]
\begin{tabular}{|l|l|l|l|l|l|}
\hline
\multirow{2}{*}{Instance Name} & \multirow{2}{*}{Description} & \multirow{2}{*}{Cost} & \multicolumn{3}{l|}{Performance}	\\ \cline{4-6} 
                               &                              &                       & $A_1$      & $A_2$      & $A_3$ 	\\ \hline
$it_1$                          & Small general type           & 5                     & 20        & 24        & 22			\\ \hline
$it_2$                          & Big general type             & 10                    & 11        & 13        & 12			\\ \hline
$it_3$                          & CPU optimised type           & 10                    & 10        & 15        & 9			\\ \hline
$it_4$                          & Memory optimised type        & 10                    & 10        & 9         & 12			\\ \hline
\end{tabular}
\caption{Costs and Performances}
\label{tab:its}
\end{table}

\subsubsection{Budget} The budget constraint was set to different values ranging from 40 to 85.

\subsection{Experimental Results}

\begin{figure}[h!]
	\centering
		\includegraphics[width=0.5\textwidth,height=0.3\textheight]{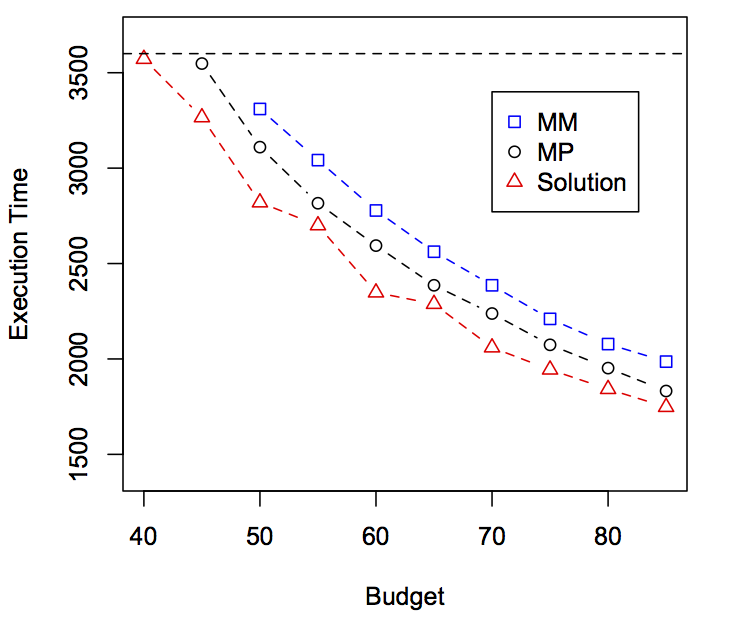}
	\caption{Execution Times for Different Approaches}
	\label{fig:result}
\end{figure}

The result is shown in Figure \ref{fig:result}. The x-axis represents the budget while the y-axis is the execution time. The black horizontal dashed line represents 3600 seconds, i.e. an hour.

It can be seen that, given the same budget, our approach, i.e. red and triangle line, always had the lower execution time in compare to other 2 simple approaches. In comparison with the MI approach, ours was able to reduce the execution time by average 13\%. The MP approach which focused on maximising the execution parallelism by choosing the cheapest instance type performed better than MI approach, which preferred more expensive instance type. However, in average, its execution times were still 7\% higher than the proposed one.

Furthermore, our approach was also able to handle the low budget constraint: while MP required the budget to be at least 45 and MI could not satisfy any budget below 50, our approach satisfies the budget as low as 40.

\begin{figure}[h!]
	\centering
		\includegraphics[width=0.5\textwidth,height=0.3\textheight]{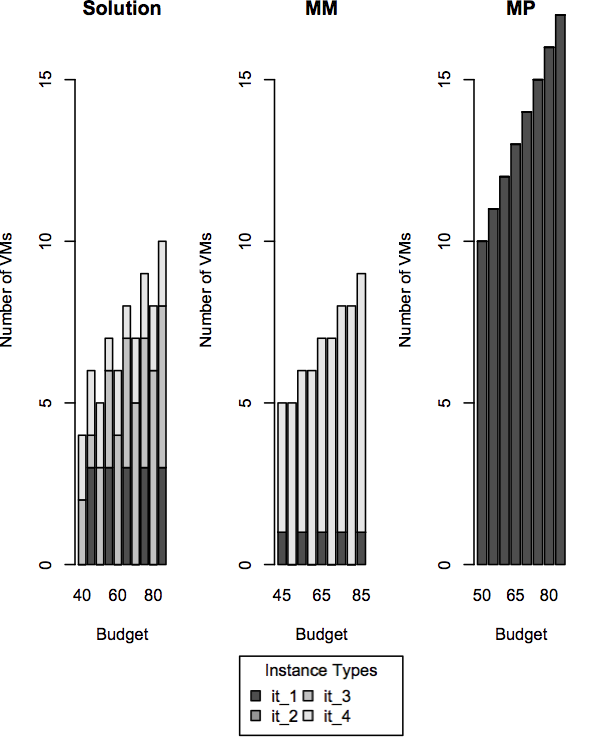}
	\caption{Number of VMs of Each Type}
	\label{fig:result_its}
\end{figure}

As mentioned earlier, there is a trade-off between minimising an individual task's execution time with maximising the parallelism. The trade-off is presented based on instance selection: powerful but expensive versus less powerful but cheap instance types. Moreover, using the combination of different instance types usually results in better performance in compare to selecting one instance type.

Figure \ref{fig:result_its} shows the number of VMs and their instance types used by different approaches for different values of the budget constraint. It can be seen that the MP (right figure) approach always went for cheapest instance type (i.e. $it_1$) and managed to maintain the highest number of VMs. On the other hand, the MI approach (middle figure) tried to use as much VMs of instance type $it_4$ as possible since it had the best average performance in compare to other three. If there were any remaining budget, MI added an additional VM of $it_1$ in order to increase the performance.


Instead of following only on trend to select instance type, our approach (left figure) was more flexible. When the budget is 40, 50, 60, 70 and 80, it prioritised execution parallelism by adding more VM of the cheapest instance type $it_1$. However, then the budget is 45, 55, 65, 77 and 85, none VM of $it_1$ was created. Instead, VMs of $it_3$ and $it_4$, which had the best performance for tasks of $A_2$ and $A_3$, were created in order to reduce the overall individual task execution time. As the result, our approach could achieve the better performance with the same budget constraint.

\section{Conclusion}
\label{conclusion}
BoT applications have been widely used in not only scientific but also industrial communities. However, they require a huge amount of resources which can only be satisfied in a distributed environment consisting of many interconnected machines. Many efforts have been spent on optimising the execution of BoT applications on grid computing in which resources are already available and users have to compete with each other to acquire free resources. Hence, the scheduling of BoT on the Grid mainly focuses on assigning tasks to the `best' suited machines.

Cloud computing on the other hand provides an isolated environment (not taking into account multitenancy)  in which a user does not need to share her resources with anyone else. Moreover, it is also possible to select the resource types which are best suited for the applications. However, cloud computing resources are not free of charge and a user has to pay as soon as the VMs start running. Hence, the problem of executing BoT applications on the cloud is not only about assigning tasks to resources but also selecting the type of resource(s), which are most appropriate. Moreover, with multiple applications to run, the problem is further complicated as each application potentially requires different types of resource for to achieve the best performance.

In this paper, we investigated the execution of multiple BoT applications on the cloud given a budget constraint. The problem is modelled and a heuristic algorithm was proposed in order to decide the selection of different cloud resources and the assignment of tasks onto resources. By comparing our approach to other simple ones, it was shown that the proposed heuristic algorithm was able to reduce the execution time from 4\% to 15\% given the budget constraint.

For future work, we plan to further expand our heuristic algorithm to take into account the execution deadline while minimising the cost. Moreover, we also want to incorporate dynamic scheduling feature to handle any unexpected issues during runtime, which are inevitable in real-time execution on the Cloud. Finally, we want to support scheduling tasks whose execution times are unknown, i.e. non-clairvoyant scheduling.

\section*{Acknowledgment}
This research is supported by the EPSRC grant `Working Together: Constraint Programming and Cloud Computing' (EP/K015745/1), a Royal Society Industry Fellowship `Bringing Science to the Cloud', an EPSRC Impact Acceleration Grant (IAA) and an Amazon Web Services (AWS) Education Research Grant.

\bibliographystyle{ieeetr}
\bibliography{references}

\end{document}